\documentclass[aps,pre,reprint,unsortedaddress,superscriptaddress]{revtex4-1} 

\usepackage{amsmath}  
\usepackage{amsfonts} 
\usepackage{graphicx} 

\usepackage{amssymb}
\usepackage{gensymb}

\usepackage{dcolumn}
\usepackage{bm}
\usepackage{subcaption} 
\DeclareCaptionJustification{justified}{\justifying}
\usepackage{verbatim}

\usepackage{color}
\usepackage{epstopdf}
\usepackage{hyperref}
\usepackage{xifthen}
\usepackage[commandnameprefix=always]{changes}
\usepackage{ulem}  
\usepackage{upgreek} 
\usepackage{xtab, afterpage}
\usepackage[version=4]{mhchem}
\usepackage[english]{babel}

\DeclareUnicodeCharacter{0308}{\"{}}

\DeclareMathAlphabet{\mathpzc}{OT1}{pzc}{m}{it}

\newlength\dlf  

\usepackage{ragged2e}

\begin{document}

\title{
Dynamic principles of concentration buffering through liquid-liquid phase separation
}%

\author{Logan de Monchaux-Irons}
\affiliation{Department of Biology, Institute of Biochemistry,
ETH Zurich, Otto Stern Weg 3, 8093, Zurich, Switzerland}
\affiliation{Bringing Materials to Life Initiative, ETH Zurich, Switzerland}
\author{T-Y Dora Tang}
\affiliation{Department of Synthetic Biology, Campus Saarbrucken, University of Saarland, Germany}
 
\author{Christoph A. Weber}
\affiliation{Faculty of Mathematics, Natural Sciences, and Materials Engineering: Institute of Physics, University of Augsburg,
Universitätsstrasse 1, 86159 Augsburg, Germany}

\author{Thomas C.T. Michaels}
\email{thomas.michaels@bc.biol.ethz.ch}
\affiliation{Department of Biology, Institute of Biochemistry,
ETH Zurich, Otto Stern Weg 3, 8093, Zurich, Switzerland}
\affiliation{Bringing Materials to Life Initiative, ETH Zurich, Switzerland}

\begin{abstract}
Living systems must maintain robust biochemical function despite fluctuations that span a wide range of timescales. Biomolecular condensates formed by liquid–liquid phase separation (LLPS) have been shown to buffer concentration fluctuations, but the principles governing their dynamic regulation remain unclear.
We address this by probing the response of LLPS to oscillatory perturbations that mimic fluctuations across different timescales, establishing the first systematic frequency-domain analysis of concentration buffering by condensates.
We find that condensates act as frequency-selective filters: the perturbed dilute phase behaves as a high-pass filter, while the dense phase attenuates both low- and high-frequency perturbations. We establish quantitative links between LLPS parameters including interaction strength, droplet size, and molecular diffusivity, and the timescale range over which condensates effectively buffer concentration fluctuations. 
These findings establish the fundamental dynamical limits of concentration buffering by LLPS, with implications for how cells may use LLPS to adapt to fluctuating environments and for the design of synthetic condensates with programmable control properties.
\\

{\bf{Significance statement:}} Cellular function depends on stability under fluctuating conditions that occur across diverse timescales. Biomolecular condensates formed by liquid–liquid phase separation have been shown to buffer concentration fluctuations. However, the dynamic response to fluctuations at different timescales remains unclear. 
Here, we use control theory, analytic modeling, and simulations to reveal the principles of dynamic concentration buffering. We show that condensates act as frequency-selective filters: they suppress slow fluctuations but may transmit fast ones. We identify key timescales and physical parameters that govern buffering effectiveness. These results establish the fundamental dynamical limits of condensate-mediated buffering and provide a general framework for understanding how cells may exploit phase separation to regulate fluctuations.

\end{abstract}

\maketitle 

\cleardoublepage

\section{\label{sec:introduction} Introduction}

Cells must function robustly despite constant fluctuations in their molecular environment. 
Gene expression noise, metabolic variability, and changing external conditions generate concentration perturbations across a wide range of timescales \cite{elowitz2002stochastic,alberts1998cell}, necessitating mechanisms to control the cellular environment \cite{batada2007evolution,stoeger2016passive}. One suggested control mechanism are biomolecular condensates formed by liquid-liquid phase separation (LLPS) \cite{banani2017biomolecular,hyman2014liquid,brangwynne2009germline,fritsch2021local}. Biomolecular condensates are involved in diverse cellular functions \cite{molliex2015phase,strom2017phase,shin2017liquid,wei2020nucleated} and have been shown, both theoretically \cite{Deviri2021,zechner2025concentration} and experimentally \cite{Klosin2020}, to buffer protein concentrations. 
In binary LLPS systems (i.e., mixtures of one solute in a solvent), concentrations are buffered as the equilibrium concentrations of components within each phase are fixed by thermodynamics, even if the total concentrations of the system change. This implies that LLPS can implement a cellular concentration control mechanism \cite{weber2021drops} when the system is in static equilibrium. However, cells experience concentration fluctuations across many timescales, yet we do not know whether condensates can buffer fast perturbations as effectively as slow ones and how the effectiveness of buffering at different timescales depends on condensate properties, such as interaction strength or size.

A natural way to study such dynamical regulation is to use the framework of control theory. 
Control theory provides powerful methods for analyzing how systems respond to perturbations and maintain stability \cite{mayr1970origins,doyle2013feedback}. 
At its core is the principle of feedback: the comparison between a desired output and the actual system state to reduce error. 
Feedback control underlies both engineered systems, from thermostats to flight guidance \cite{astrom95,doyle2013feedback,EDMUNDS1979}, and natural systems, including bacterial chemotaxis \cite{yi2000robust}, calcium homeostasis in mammals \cite{ELSAMAD200217}, and the resilience of insect societies \cite{schmickl2018integral}. 
While control-theoretic analysis has been applied to homogeneous biochemical networks \cite{behre2008structural,lee2003roles,huang1996ultrasensitivity}, its extension to spatially heterogeneous systems such as biomolecular condensates has remained largely unexplored \cite{weber2021drops}. 

Here, we apply control theory to condensates to uncover the dynamic principles of concentration buffering. 
Using analytic modeling and simulations, we perform the first systematic frequency-domain analysis of LLPS. 
We show that condensates act as frequency-selective filters, suppressing some frequencies of concentration fluctuations while transmitting others. 
By linking filtering behavior to LLPS parameters such as interaction strength, droplet size, and molecular diffusivity, we identify the fundamental dynamical limits of condensate-mediated buffering. 
These results suggest a mechanism for how phase separation could contribute to cellular robustness and offer design principles for engineering synthetic condensates with tunable control properties. 

\section{Results}

\subsection*{Control theory}
Control strategies can be broadly classified into two categories: open-loop control and closed-loop control \cite{bode1945network,aastrom2021feedback}. Open-loop control employs static parameters optimized for specific conditions to minimize error, but it lacks adaptability to dynamic changes. In contrast, closed-loop control leverages feedback: it continuously measures the system error and adjusts inputs dynamically to bring the output closer to the set-point. For example, bacterial chemotaxis employs closed-loop control to adapt to changing concentrations of chemoattractants, ensuring robust directional movement \cite{barkai1997robustness, yi2000robust}.

\begin{figure*}
    \centering
    \includegraphics[width=1\linewidth]{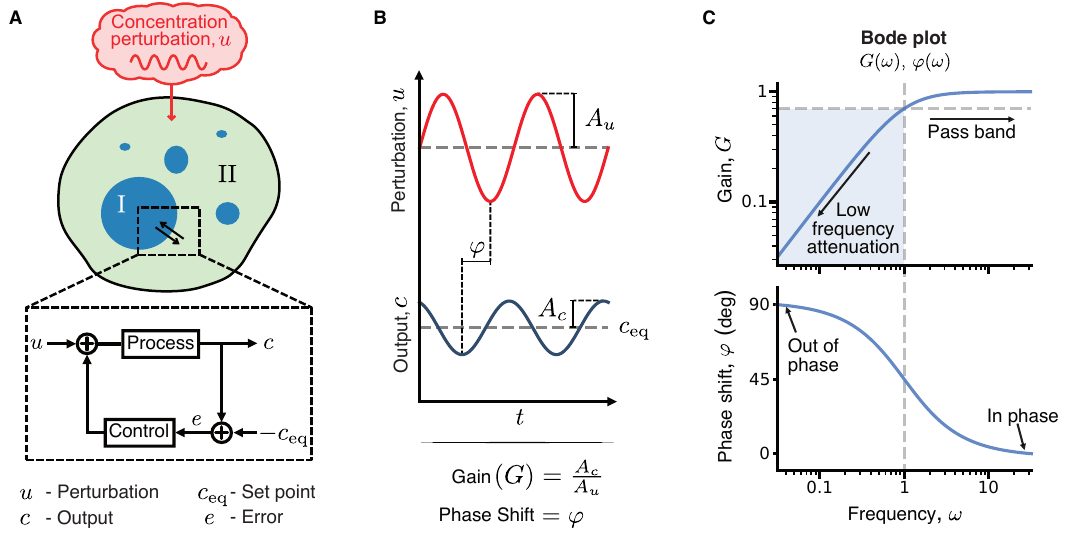}
    \caption{\textbf{Schematics of LLPS-based feedback control and frequency response characterization.} \textbf{A}  Schematic of the LLPS feedback control system. A process produces an output concentration $c$ in phase I from an input concentration perturbation $u$ in phase II. We then measure the error,    $e=c_{\rm{eq}}-c$, where $c_{\rm{eq}}$ is the desired concentration in phase I. The error is then operated on by a controllers and added back to the input. \textbf{B} Schematic of a sinusoidal input $u$ and the output $c$ showing how the gain $G$ and phase shift $\varphi$ are measured. \textbf{C} Bode plots are constructed by varying the input frequency and measuring $G$ and $\varphi$ as a function of $\omega$. This schematic shows the behavior of a high-pass filter that attenuates low frequency perturbations and allows high frequencies to pass through.}
    \label{fig:1}
\end{figure*}

Closed-loop control operates by tracking the system error $e=y_0-y$, i.e., the difference between current system output $y$ and desired output $y_0$ (set point). The error is then processed through a controller that uses this information to adjust inputs and correct the error (Fig.~\ref{fig:1}(a)). The controller can perform various operations on the error to steer the system under control, for example, proportional-integral (PI) control \cite{aastrom1995controllers,aastrom2000model}.

The characteristics of a control system are commonly analyzed using the frequency response of the system to oscillatory input perturbations with a form $u(t)=A_u\sin(\omega t)$ of varying frequency $\omega$. For a linear system, the resulting output will generally be oscillatory with potentially altered
amplitude and phase. The ratio of the output amplitude $A_y$ to the perturbation amplitude $A_u$ is called
the gain $G=|A_y/A_u|$, which indicates how the system attenuates perturbations (Fig.~\ref{fig:1}(b)). The phase shift $\varphi$ measures the delay between the input and output signals. By varying the input frequency
$\omega$ and measuring the gain $G(\omega)$ and the phase shift $\varphi(\omega)$, we can construct the full frequency response of the system, the so-called Bode plot \cite{bode1945network,ingalls2004frequency,csete2002reverse} see Fig.~\ref{fig:1}(c) for an illustration.
Bode plots characterize the performance of a feedback system, including filtering effects. Feedback systems can attenuate some frequencies of perturbations and allow others to pass through creating a frequency filter. Bode plots visualize this filter effect by showing lower gain for attenuated frequencies and high gain for frequencies that pass through. For example, Fig.~\ref{fig:1}(c) depicts the Bode plot of a high-pass filter where low frequency perturbations are attenuated but high frequencies can pass through.

In this paper, we perform the first Bode plot analysis of LLPS concentration buffering to assess its perturbation rejection capability as a function of the frequency $\omega$ of the disturbance. To this end, we build on theoretical work \cite{bauermann2022chemical, weber2021drops,zechner2025concentration} to devise a simple model of a dynamic LLPS system and employ control theory analysis tools to characterize its control characteristics.

\subsection*{Model of phase-separation dynamics}

We consider a system of fixed, incompressible total volume $V$, consisting of two coexisting phases: a dense phase (phase I) with volume $V^{\rm{I}}$, which is enriched in a biomolecular component (e.g., a protein A), and a surrounding dilute phase (phase II), which contains a lower concentration of A and is rich in another component (e.g., solvent B), see Fig.~\ref{fig:1}(a). The exchange of A between phases occurs through molecular diffusion, driven by differences in chemical potential, ensuring that the system maintains phase equilibrium. The total amount of A in the system can fluctuate due to synthesis, degradation, or external perturbations, but phase separation dynamically redistributes A between the two phases, allowing the system to buffer changes in concentration after a perturbation.

To describe the system's dynamics, we formulate a set of coupled ordinary differential equations (ODEs) governing the concentrations of A in each phase, $c^{{\rm{I}}}$ and $c^{{\rm{II}}}$, as well as the condensate volume $V^{\rm{I}}$:
\begin{subequations}\label{eq1}
\begin{align}
    \frac{{\rm{d}}c^{\rm{I}}}{{\rm{d}}t} &= \frac{J^{\rm{I}}}{V^{\rm{I}}} -\frac{c^{\rm{I}}}{V^{\rm{I}}}\frac{{\rm{d}}V^{{\rm{I}}}}{{\rm{d}}t}\,, \\
    \frac{{\rm{d}}c^{\rm{II}}}{{\rm{d}}t} & =-\frac{J^{\rm{I}}}{V^{{\rm{II}}}}\, +\frac{c^{{\rm{II}}}}{V^{{\rm{II}}}}\frac{{\rm{d}}V^{{\rm{I}}}}{{\rm{d}}t}+\frac{{\rm{d}}u}{{\rm{d}}t}\,,
\end{align}
\end{subequations}
where $V^{\rm{II}}=V-V^{\rm{I}}$ and $\text{d} u/ \text{d} t$ is an external perturbation term to capture fluctuations in total protein concentration and examine the system's response to such perturbations, which conserves the fixed volume ($u_{\rm{solute}}=-u_{\rm{solvent}}$).  The full mathematical details with explicit equations can be found in the SI. These dynamics are influenced by the flux of A between phases, denoted $J^{\rm{I}}$, which depends on the chemical potential difference that drives phase equilibrium \cite{bauermann2022chemical,laha2024chemical}:
\begin{align}
    J^{\rm{I}}=L_0(\mu^{\rm{I}}-\mu^{\rm{II}})\, ,
\end{align}
where $L_0$ is a mobility coefficient that we set to a constant, $L_0$ sets the timescale for the effective diffusivity between phases, $D_e=k_BTL_0$. $\mu =\partial f/\partial c$ is the chemical potential, obtained from the free energy density of the system, $f$. Therefore, to obtain $\mu$, we need an explicit free-energy model for LLPS. To this end, we employ here the Flory–Huggins model, a well-established thermodynamic framework for phase separation \cite{flory1942thermodynamics,huggins1942some,weber2019physics}. We note that our results are qualitatively robust to the choice of $f$ and thus other LLPS models can be used for this control analysis \cite{overbeek1957phase,Celora2023}. In the Flory–Huggins model, the free energy density is given by
\begin{equation}\label{eqs1}
   f=\frac{k_B T}{\nu_0}\left[\phi\ln\phi +(1-\phi)\ln(1-\phi)+\chi\, \phi\, (1-\phi)
    \right ],
\end{equation}
where $\nu_0$ is the molecular volume of A and $\phi = \nu_0 c$ is the molecular volume fraction of component A.
In this model, phase separation arises from the interplay between enthalpic interactions, which favor demixing, and entropic contributions, which promote mixing. This balance is quantified by the Flory interaction parameter $\chi$, which controls the degree of phase separation and dictates the equilibrium concentrations in each phase.

\subsection*{Frequency analysis and Bode plots for condensates}

\begin{figure*}[ht!]
    \centering
    \includegraphics[width=\linewidth]{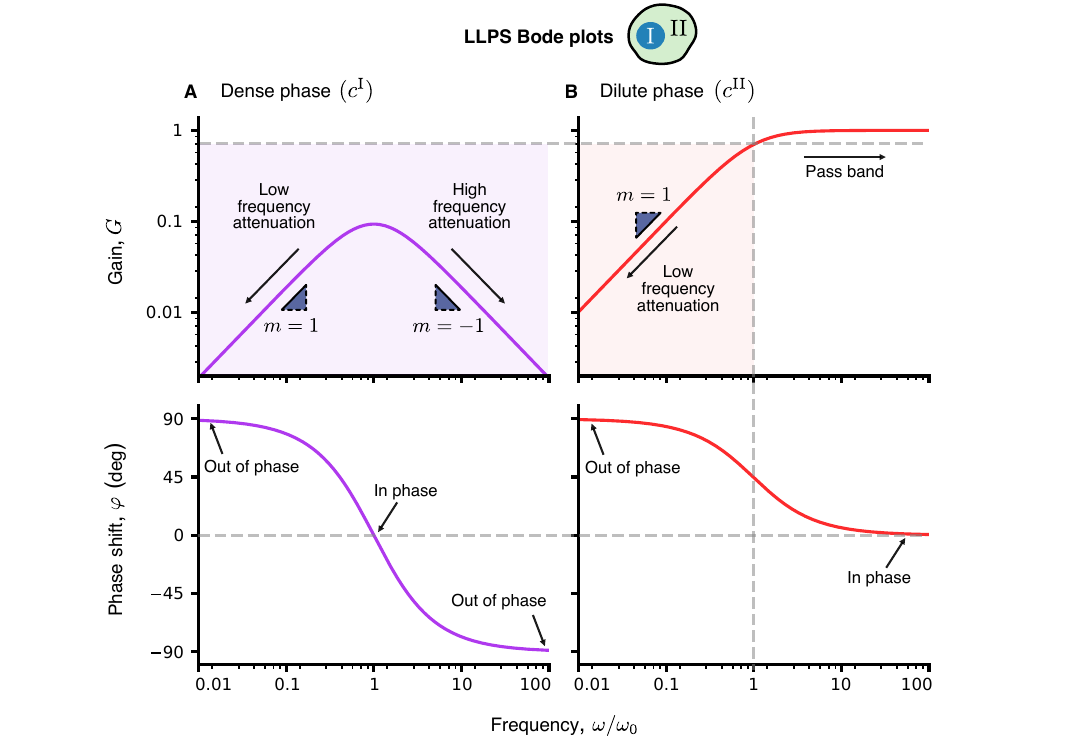}
    \caption{\textbf{Bode plots showing the frequency response of our LLPS system to concentration perturbations in the dilute phase.}  Frequency-domain analysis shows distinct filtering properties in each phase: the dense phase (\textbf{A}) attenuates both slow and fast fluctuations, whereas the dilute phase (\textbf{B}) acts as a high-pass filter, transmitting fast perturbations while suppressing slow ones . The LLPS parameters for both plots were $\chi = 2.09507$, $D_e= 1$ and $V^{\rm{I}}=0.5$. The frequency is expressed in units of $\omega/\omega_0$ where $\omega_0=D_e$}
    \label{fig:3}
\end{figure*}

Due to the non-linearity of the dynamic equations Eq. \eqref{eq1}, we first performed the frequency analysis numerically. We add perturbations of the type $\frac{\text{d}u}{\text{d}t} = A_u\omega \cos (\omega t)$ to the concentration dynamics of phase II and numerically solve the system over a wide range of frequencies $\omega$. For small perturbations, the system behaves linearly from equilibrium, so we select $A_u$ as 5$\%$ of the equilibrium dilute phase concentration. We note that the derivative of the perturbation $u$ is added such that after numerical integration the resulting perturbation amplitude does not depend on $\omega$ to stay in the small perturbation regime.  We use the fast Fourier transform (FFT) to convert to the complex frequency domain, and then the argument and angle of the concentrations and volume are taken at the frequency of the perturbation. The gain is obtained by dividing the modulus of the output by the amplitude of the input perturbation $G= |A_y/A_u|$, and the phase shift, $\varphi$, is the argument of the output. 

To gain further insight into how phase separation governs feedback control, we derive explicit expressions for the transfer functions that characterize the condensate's response. By linearizing the kinetic equations \eqref{eq1} around phase equilibrium, we obtain a tractable analytical framework that allows us to connect the frequency response parameters directly to phase separation parameters.
For linear systems, the frequency domain is accessed by taking the Laplace transform of the system \cite{bode1945network}. To analyze how inputs affect the system outputs, we divide the transformed output by the transformed input to obtain the output-to-disturbance transfer function, which is defined below (see Eq.~\eqref{eqHod}). The gain and phase shift can then be calculated as the modulus and argument of the transfer functions, respectively.
We linearize the system outputs $\boldsymbol{y}$ around the set points $\boldsymbol{r}$ as follows
\begin{subequations}   \label{ode1}
\begin{equation}
    \boldsymbol{\dot{y}}= \boldsymbol{\Lambda}|_{\rm{eq}}\,(\boldsymbol{r}-\boldsymbol{y})+\boldsymbol{\dot{u}}
\end{equation}
where
\begin{equation}
    \boldsymbol{y}=\begin{pmatrix}
           c^{\rm{I}} \\
           c^{\rm{II}} \\
         \end{pmatrix},\,
         \boldsymbol{r}=\begin{pmatrix}
           c^{\rm{I}}_{\rm{eq}} \\[2pt]
           c^{\rm{II}}_{\rm{eq}} \\
         \end{pmatrix},\,
         \boldsymbol{u}=\begin{pmatrix}
           0 \\
           u \\
         \end{pmatrix}
\end{equation}
\end{subequations}
and $\boldsymbol{\Lambda}|_{\rm{eq}}$ is the Jacobian matrix of the system Eq. \eqref{eq1} evaluated at phase equilibrium. Taking the Laplace transform of the system of equations Eq. \eqref{ode1} to access the frequency domain yields:
\begin{equation}\label{ode2}      \boldsymbol{\bar{y}}=\boldsymbol{PC}(\boldsymbol{r}-\boldsymbol{\bar{y}})+\boldsymbol{P\bar{u}}\,,
\end{equation}
where we used the fact that the set point $\boldsymbol{r}$ is constant, $\boldsymbol{P}$ and $\boldsymbol{C}$ are matrices and $\bar{x}(s) = \int_0^\infty x(t) e^{-st}dt$ denotes the Laplace transform of $x$. Solving Eq.~\eqref{ode2} for $\boldsymbol{\bar{y}}$ yields:
\begin{align}\label{ode3}
    \boldsymbol{\bar{y}}=&(\boldsymbol{I}+\boldsymbol{PC})^{-1}\boldsymbol{PC}\,\boldsymbol{r}+(\boldsymbol{I}+\boldsymbol{PC})^{-1}\boldsymbol{P\,\bar{u}}\, .
\end{align}
Eq.~\eqref{ode3} has two terms: the first term, proportional to the set point $\boldsymbol{r}$, describes the response to changes in the desired output, whereas the second term, proportional to the disturbance $\boldsymbol{u}$, captures the dynamic response to perturbations. We are interested in how the output $\boldsymbol{\bar{y}}$ relates to the inputted perturbations $\boldsymbol{\bar{u}}$, so we consider the matrix 
\begin{equation}\label{eqHod}
   \boldsymbol{H}_{\rm{od}}=(\boldsymbol{I}+\boldsymbol{PC})^{-1}\boldsymbol{P} \, ,
\end{equation} 
which is the matrix of the output-to-disturbance (od) transfer function \cite{aastrom1995controllers}. Since for our system $\boldsymbol{u}$ has only one nonzero element, there are only two elements in $\boldsymbol{H}_{\rm{od}} $ to consider.
We find that the output-to-disturbance transfer function for concentration in each phase is 
\begin{align}
    H_{\rm{od}}^{\rm{I}}(s) &= \frac{K_g s}{K_d s^2 +K_ps + K_i}\, , \label{eqan1} \\
    H_{\rm{od}}^{\rm{II}}(s) &= \frac{K_ds^2+K_ls}{K_d s^2+K_ps +K_i}\, , \label{eqan2}
\end{align}
 where $K_p$, $K_d$, $K_i$, $K_g$ and $K_l$ depend on the equilibrium concentrations and volume (explicit expressions for the parameters are given in the SI). The resulting transfer functions provide a link between the LLPS parameters and the regulatory properties of the condensate. 
To analyze the system's frequency response, we construct Bode plots by setting $s=i\omega$ and evaluating the gain as $ G = \text{abs}(H_{\rm{od}}) $ and the phase shift as $ \varphi = \arg(H_{\rm{od}}) $, where $ \text{abs}(\cdot)$ and $\arg(\cdot)$ denote the magnitude and argument (i.e., phase angle) of a complex number, respectively.

The resulting Bode plots for $c^{\rm{I}}$ and $c^{\rm{II}}$ computed using the analytical expressions of the transfer functions are shown in Fig. \ref{fig:3}. The analytical Bode plot and the numerical Bode plot methods are in agreement (see SI Fig. \ref{fig:s1}).  We find a variety of perturbation responses within the LLPS model. In the perturbed phase (phase II), we see that protein concentration $c^{\rm{II}}$ responds with low gain at low frequencies, which then plateaus to a gain of 1 at around a characteristic frequency of $\omega_0 = D_e$, where $D_e$ is an effective diffusion constant between the phases. The volume of the phases exhibits a gain near 1 at low frequencies, with a switch around $\omega_0$ and then attenuates high-frequency perturbations. Strikingly, the protein concentration in phase I attenuates both high and low frequency perturbations, with the peak gain being an order lower than the peak gain of $c^{\rm{II}}$.

A prediction from our model is that low-frequency attenuation emerges on timescales longer than the interphase exchange time $\tau_{\mathrm{e}}$ set by the effective inter-phase diffusivity $D_e$. Reported exchange measurements for in vitro FUS droplets place $\tau_{\mathrm{e}}$ on the order of seconds \cite{novakovic2025llps}. Using our Bode plot framework, we would expect the low-frequency concentration attenuation of the FUS droplets to be on the order of minutes$^{-1}$ or longer and high-frequency attenuation at seconds$^{-1}$ or shorter. This implies a characteristic cutoff frequency $\omega_0\sim D_e$: perturbations with $\omega\ll \omega_0$ are strongly buffered, whereas faster inputs ($\omega\gtrsim \omega_0$) are increasingly transmitted in the dilute phase. This mapping is consistent with experiments demonstrating concentration buffering at (quasi)steady state \cite{Klosin2020}, a regime corresponding to $\omega\to 0$ in our theory. Conversely, our analysis predicts reduced buffering efficacy for rapid perturbations on second-scale times (e.g.\ fast input pulses or shocks), particularly when exchange is slowed by interfacial barriers \cite{zhang2024exchange} or when droplets are small, which together shift $\omega_0$ to lower values. Because condensate size, material properties, and exchange kinetics vary widely across systems \cite{taylor2019quantifying,zhang2024exchange,novakovic2025llps}, the frequency window over which buffering is effective is expected to be system specific and tunable.

\begin{figure*}[t!]
    \centering
    \includegraphics[width=\linewidth]{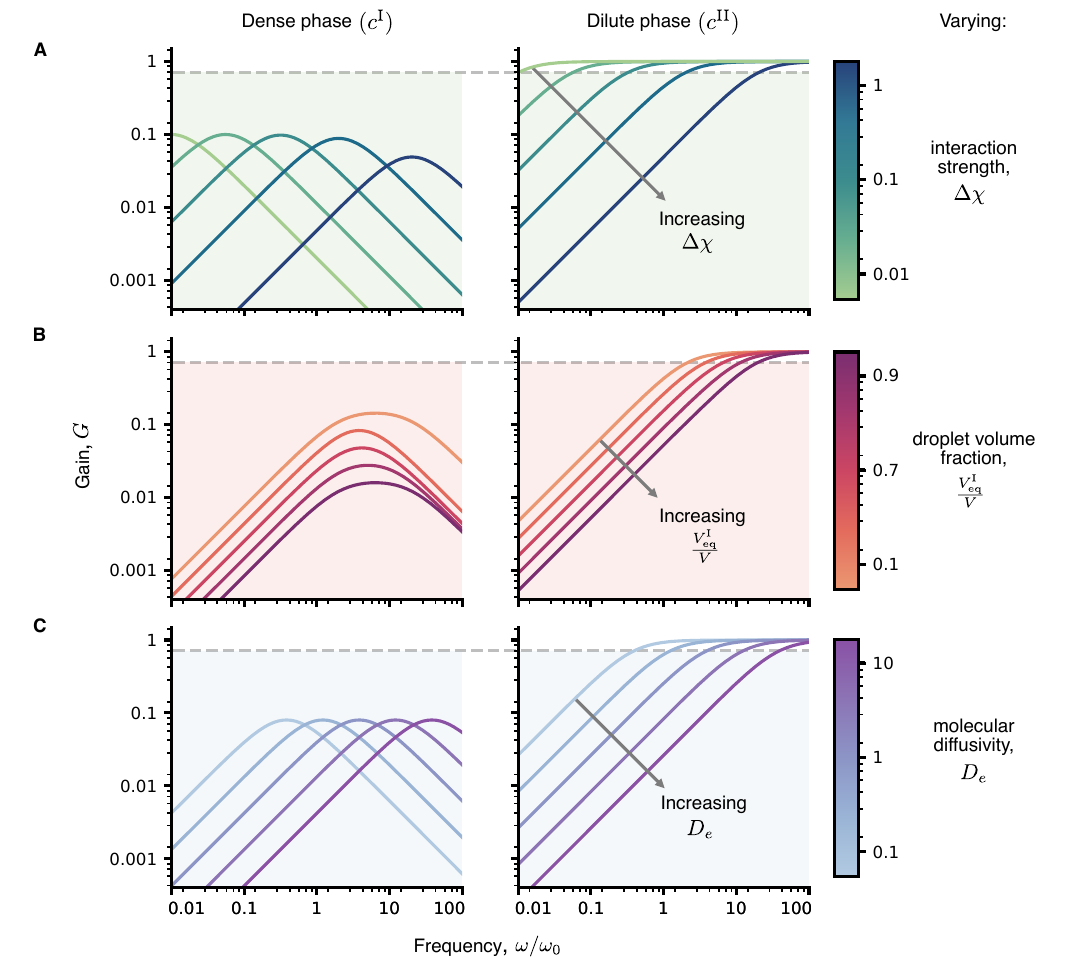}
    \caption{{\bf{The frequency response of the output to disturbance gain under different LLPS parameters.}} The parameters tested were \textbf{A} the interaction parameter $\chi$ above the critical point ($\Delta\chi=\chi-2$), \textbf{B} the equilibrium volume fraction of the droplet phase, $V^{\rm{I}}_{\rm{eq}}/V$, and \textbf{C} the effective diffusion constant $D_e$. Condensate properties tune the frequency range of effective buffering. Stronger interactions, larger droplet volumes, and faster diffusion broaden the regime of attenuation, while weak interactions or small condensates reduce buffering efficiency. Parameters for all plots are $\Delta\chi=0.3105$, $V^{\rm{I}}_{\rm{eq}}/V=0.5$, $D_e = 1$ except when a specific parameter is changed. In increasing order: (\textbf{A}) $\Delta\chi=0.001,\,0.00562,\,0.0316,\,0.178,\,1.0$; (\textbf{B}) $V^{\rm{I}}_{\rm{eq}}/V=0.1,\,0.48,\,0.7,\,0.83,\,0.9$; (\textbf{C}) $D_e=0.1,\,0.316,\,1.0,\,3.16,\,10.0$. }
    
    \label{fig:4}
\end{figure*}

\subsection*{Connecting LLPS parameters and control properties}

By embedding our model of phase separation into a control-theoretical framework, we establish a direct quantitative connection between the thermodynamic and kinetic parameters of LLPS and the disturbance rejection properties of biomolecular condensates. Specifically, we use Bode plots (Fig.~\ref{fig:4}) to explore how key parameters, including interaction strength, equilibrium volume, and diffusivity, shape the frequency response of the system.

We first examine the role of the interaction parameter $\chi$, which quantifies the drive to demix competing with the mixing entropy (Fig.~\ref{fig:4}a). Decreasing the interaction strength $\chi$ weakens the phase separation. As a result, the gain curve shifts to the left, effectively reducing the system’s ability to attenuate low-frequency perturbations. As we approach the critical interaction parameter (i.e., $\Delta\chi = \chi - \chi_c \to 0$), the distinct two-phase regime disappears, and so does the low-frequency filtering. This finding shows that phase separation is necessary for low-frequency attenuation. In biological terms, this implies that proximity to the critical point limits a condensate’s ability to buffer against long-timescale changes in protein concentration.

Next, we analyze the effect of the equilibrium volume fraction of the droplet phase (Fig.~\ref{fig:4}b). Larger droplets (i.e., bigger phase volume fraction $V^{\rm I}_{\rm eq}/V$) lead to lower maximum gain in the dense phase. Physically, this behavior reflects the increased capacity of larger droplets to absorb and redistribute excess protein, thus dampening the system response amplitude. At the same time, increasing the droplet volume also improves low-frequency attenuation in the dilute phase. Enhanced attenuation arises from a larger dense phase acting as a more effective sink, enabling slower redistribution of solute molecules over longer timescales, thus enhancing the system’s buffering capability.

Lastly, we investigate the impact of the effective molecular diffusivity through the combined parameter $D_e$, which governs the rate at which material is exchanged between phases (Fig.~\ref{fig:4}c). Increasing $D_e$ shifts the frequency response curve to the right, improving the attenuation of low-frequency perturbations while allowing faster responses at higher frequencies. This reflects the fact that faster diffusion enables more efficient equilibration between the two phases. From a biological perspective, this implies that LLPS systems with higher diffusivity may be better equipped to reject slow fluctuations, thereby enhancing robustness under noisy intracellular conditions.

Taken together, these results suggest that LLPS may be actively tuned to implement frequency-selective control. Modulating interaction strength, droplet volume, and molecular diffusion could dynamically regulate the responsiveness of condensates to external fluctuations. Our analysis provides a theoretical foundation for understanding such control strategies and sets the stage for future experimental investigations into the tunable regulatory roles of phase-separated compartments.

\section{Discussion}

Our results show that concentration buffering by biomolecular condensates is inherently dynamic and constrained by timescales. While condensates are often described as stabilizing intracellular environments, we show that this stabilization depends critically on the relative frequency of fluctuations and on condensate properties such as size, interaction strength, and diffusivity. Recognizing these dynamical limits provides a new perspective on condensate function in cells, where fluctuations span timescales from seconds to hours, and suggests that phase separation may be tuned to filter specific ranges of perturbations rather than universally filter fluctuations.

Our analysis reveals that condensates exhibit distinct frequency-dependent filtering properties in each coexisting phase. In the dilute phase, where the perturbation is introduced, we observe a high-pass filter behavior: low-frequency perturbations are strongly attenuated, while high-frequency perturbations are transmitted. In contrast, the dense phase behaves as a band-stop filter, attenuating both low- and high-frequency perturbations. This suggests that the phase boundary itself contributes to an effective filtering mechanism, potentially adding an additional layer of regulation to the dynamics of macromolecular concentrations.

Importantly, we find that the filtering characteristics are tunable and depend on key thermodynamic parameters of LLPS. When the interaction strength $\chi$ approaches the critical value $\chi_c$, the condensate loses its ability to attenuate low-frequency perturbations. Conversely, strong interactions enhance low-frequency disturbance rejection, suggesting that phase separation is key to filtering slow fluctuations. In biological systems, interaction strength can be modulated by mutations or post-translational modifications that alter multivalent binding affinity, as shown for RNA-binding proteins such as FUS and other prion-like domain–containing proteins \cite{banani2017biomolecular,molliex2015phase, rhoads2018role, ruff2021ligand}. Such tunability provides a mechanism by which cells could adapt the frequency range of concentration buffering to distinct physiological conditions.

In addition to interaction strength, other condensate properties such as droplet size and molecular diffusivity also modulate dynamic buffering. Larger droplets filter more effectively at low frequencies because of reduced exchange rates with the dilute phase, whereas smaller droplets transmit fluctuations more readily. In cells, condensate size is highly variable and can be regulated by protein concentration, nucleation kinetics, or coalescence dynamics \cite{banani2017biomolecular,hyman2014liquid,lafontaine2021nucleolus}. Similarly, the diffusivity of condensates sets the timescale of equilibration: slower diffusion enhances low-frequency rejection but reduces responsiveness to rapid changes. Diffusivity is itself tunable, for example, through changes in crowding or mutations that alter protein disorder \cite{kaur2019molecular,novakovic2025llps}.

Bode plots, widely used in engineering and physical sciences, provide an intuitive and quantitative method to characterize such control properties. They are broadly applied to physical systems, from flight control systems \cite{EDMUNDS1979} and how pilots interact with them \cite{Hess2006} to battery design \cite{jang2011equivalent,bodebattery2019}, and are increasingly being applied in biology, including neural dynamics \cite{smith2022stimulating,van2013modeling}, signaling cascades \cite{mettetal2008frequency} and synthetic gene circuits \cite{dolan2012loop,teo2019artificial}. Our work extends this methodology to LLPS, offering a framework to probe the dynamic regulation of biomolecular condensates. We anticipate that this type of control-theoretical analysis will be broadly applicable to study the dynamics of other LLPS-driven systems, including stress granules \cite{buchan2014mrnp,yang2020g3bp1}, nucleolar dynamics \cite{lafontaine2021nucleolus}, and transcriptional condensates \cite{henninger2021rna}.

Finally, to experimentally validate our predictions, we propose an in vitro assay using a well-characterized phase-separating protein such as G3BP1 or LAF-1 \cite{yang2020g3bp1,Schuster2018}. Fluorescently tagging the protein with a reporter protein can provide estimates of the protein concentration within the droplet by fluorescence methods.  A microfluidic device can couple direct imaging of a droplet with controlled input of protein concentration so that a sinusoidal concentration profile can flow over a constrained droplet \cite{Erkamp2023}, analogous to the theoretical method of Bode plot construction. The construction of experimental Bode plots would offer a direct test of the dynamic filtering properties of condensates and reveal how physical parameters shape their control function in biological contexts.

\section{Acknowledgments}
We thank Christoph Zechner and Anne-Lena Moor for useful discussions. This work was supported by ETH Zurich (L.D.M.I and T.C.T.M.), the Swiss NSF (Grant No. SNSF 219703 to T.C.T.M.).
\bibliography{scibib}

\newpage
\clearpage
\newpage

\widetext
\setcounter{equation}{0}
\setcounter{figure}{0}
\setcounter{table}{0}
\setcounter{page}{1}
\renewcommand{\theequation}{S\arabic{equation}}
\renewcommand{\thefigure}{S\arabic{figure}}
\renewcommand{\bibnumfmt}[1]{[S#1]}

\section*{Supplementary Materials}

\subsection*{Dynamics of Two-component LLPS}
We take a simple system with no spatial dependence with fixed total volume $V = \nu_0(N_1+N_2)$ where $\nu_0$ is the molecular volume of molecules 1 and 2. This volume is split into two regions $I/II$ with volumes $V^{\rm{I}}, V^{\rm{II}}$ and molecule numbers $N^{\rm{I}}_{i}, N^{\rm{II}}_{i}$ with $i=1,2$.

To account for exchange between the two phases, we used the linear response in which the flux of the extensive property is proportional to the affinity of its intensive property \cite{PhysRev.37.405}.
\begin{equation}
    J_k^{\rm{I}}=\sum_j L_{kj} (F^{\rm{I}}_j-F^{\rm{II}}_j)\,.
\end{equation}
In this case, the extensive properties are the number of molecules $N_1$, $N_2$ with corresponding intensive properties being their chemical potentials, $\mu_i^{I/II}$ so the associated fluxes are written as
\begin{equation}
   J_{N_1}^{\rm{I}} = L_{11}  \left(\mu
   ^{\rm{I}}_1-\mu ^{\rm{II}}_1\right)+L_{12} \left(\mu
   ^{\rm{I}}_2-\mu ^{\rm{II}}_2\right) = -J_{N_1}^{\rm{II}}\,,
\end{equation}
\begin{equation}
       J_{N_2}^{\rm{I}} = L_{21}  \left(\mu
   ^{\rm{I}}_1-\mu ^{\rm{II}}_1\right)+L_{22} \left(\mu
   ^{\rm{I}}_2-\mu ^{\rm{II}}_2\right) = -J_{N_2}^{\rm{II}}\,.
\end{equation}
The mobility coefficients $L_{ik}$ were chosen as $L_{ii} = D_0$ and $L_{ik} = 0$. This gives the dynamics of $N_i$ as 
\begin{align}
     \frac{dN_i^{\rm{I}}}{dt} &= -J^{\rm{I}}_{N_i} \\&= -\frac{dN_i^{\rm{II}}}{dt}\nonumber\,.
\end{align}
So, the dynamics of the molecular volume fractions $\phi_i^{\rm{I}}=\nu_0 N_i^{\rm{I}}/{V^{\rm{I}}}$ follows
\begin{align}
    \frac{d\phi_i^{\rm{I}}}{dt} &= \nu_0 \frac{d}{dt}\left(\frac{N_i^{\rm{I}}}{V^{\rm{I}}}\right)\\
    &=\frac{\nu_0}{V^{\rm{I}}}\frac{d{N}_i^{\rm{I}}}{dt}-\frac{\phi_i^{\rm{I}}}{V^{\rm{I}}}\frac{d{V}^{\rm{I}}}{dt}\,.
\end{align}
This is the form of the equations used in the main text (Eq. \ref{eq1}), with $V^{\rm{I}} = \nu_0(N_1^{\rm{I}} + N_2^{\rm{I}})$ and $\frac{d{V}^{\rm{I}}}{dt} = \nu_0(\frac{d{N}_1^{\rm{I}}}{dt} + \frac{d{N}_2^{\rm{I}}}{dt})$.
In order to have a closed system we need expressions for the chemical potential, to this end we use the Flory-Huggins model for simple LLPS which has a reduced free energy \cite{flory1953principles}:
\begin{equation}\label{eqs1}
    \frac{F}{V}=f=\frac{k_B T}{\nu_0}\left(\phi_1\ln\phi_1 +\phi_2\ln\phi_2+\phi_1\phi_2\chi
    \right )
\end{equation}
where $\phi_{i} = \nu_0 c_{i}$ is the molecular volume fraction of component $i=1,2$. Interactions between $1$ and $2$ are accounted for via the Flory-Huggins parameter $\chi$. From this we calculate the chemical potentials as
\begin{align}
    \mu_1 &= k_BT(1+\ln \phi_1+\chi \phi_2)\,,\\
    \mu_2 &= k_BT(1+\ln \phi_2+\chi \phi_1)\,.
\end{align}
Using the molecular volume fractions and volume of phase $I$ we have a five-component system of ODEs, explicitly:
\begin{align}
    \frac{d\phi_1^{\rm{I}}}{dt}&=\frac{D_ev_0}{V^{\rm{I}}}\left((1-\phi_1^{\rm{I}})(\mu
   ^{\rm{II}}_1-\mu ^{\rm{I}}_1)-\phi_1^{\rm{I}}(\mu
   ^{\rm{II}}_2-\mu ^{\rm{I}}_2)\right)\,,\\
   \frac{d\phi_2^{\rm{I}}}{dt}&=\frac{D_ev_0}{V^{\rm{I}}}\left((1-\phi_2^{\rm{I}})(\mu
   ^{\rm{II}}_2-\mu ^{\rm{I}}_2)-\phi_2^{\rm{I}}(\mu
   ^{\rm{II}}_1-\mu ^{\rm{I}}_1)\right)\,,\\
   \frac{d\phi_1^{\rm{II}}}{dt}&=\frac{D_ev_0}{1-V^{\rm{I}}}\left(-(1-\phi_1^{\rm{II}})(\mu
   ^{\rm{II}}_1-\mu ^{\rm{I}}_1)+\phi_1^{\rm{II}}(\mu
   ^{\rm{II}}_2-\mu ^{\rm{I}}_2)\right)+\frac{du}{dt}\,,\\
   \frac{d\phi_2^{\rm{II}}}{dt}&=\frac{D_ev_0}{1-V^{\rm{I}}}\left(-(1-\phi_2^{\rm{II}})(\mu
   ^{\rm{II}}_2-\mu ^{\rm{I}}_2)+\phi_2^{\rm{II}}(\mu
   ^{\rm{II}}_1-\mu ^{\rm{I}}_1)\right)-\frac{du}{dt}\,,\\
   \frac{d{V}^{\rm{I}}}{dt}&=D_ev_0\left((\mu
   ^{\rm{II}}_2-\mu ^{\rm{I}}_2)+(\mu
   ^{\rm{II}}_1-\mu ^{\rm{I}}_1)\right)\,,
\end{align}
$D_e=D_0 k_B T$ is the effective diffusion constant between the phases and $u$ is the perturbation in phase $II$. The perturbation $u$ is added as a derivative so that the resulting concentration profile of the perturbation is sinusoidal with an amplitude not dependent on frequency. This keeps the condition that the concentration perturbations are not far from the equilibrium value over all frequencies. We refer to the molecular volume fraction as concentration in the main text and throughout the following SI. For our numerical and analytical results $v_0=1$, so the molecular volume fraction and concentration are equivalent in these cases. 

\subsubsection*{Convergence of numerical solutions}
For a symmetric two-component LLPS system, we can make a component implicit by $\phi_1+\phi_2=1$ so $\phi_1=\phi$ and $\phi_2=1-\phi $. With the new relations, we adapt the Flory-Huggins free energy density and chemical potential.
\begin{align}
    f&= \frac{k_B T}{\nu_0}\left(\phi\ln\phi +(1-\phi)\ln(1-\phi)+\phi(1-\phi)\chi\right)\,,\\
    \mu&=(1-2\phi)\chi+\ln(\phi)-\ln(1-\phi)\,.
\end{align}
For this system, the condition for phase equilibrium is $\mu =0$. Given this, we can get a function for $\chi$ in terms of the equilibrium $\phi$
\begin{equation}
    \chi = \frac{\ln(\phi_{\rm{eq}})-\ln(1-\phi_{\rm{eq}})}{(2\phi_{\rm{eq}}-1)}\,.
\end{equation}
For the dynamics in Fig. \ref{fig:1}, we use this relation to get the exact value for $\chi_{0.8}=2.31049...$ corresponding to an equilibrium concentration $\phi_{\rm{eq}}=0.8$. To test for convergence, we start $\phi_1^{\rm{I}}=0.7$ with $\chi_{0.8}$ for an equilibrium concentration of $0.8$, and $D_e=1$ and run for 10000 steps with a timestep of 0.01. The final error between the simulated $\phi_1^{\rm{I}}$ and 0.8 was on the order $10^-{9}$.

\subsection*{Linearization and Control Analysis}

We linearize our system around phase equilibrium concentrations ($c_{\rm{eq}}$) via a first-order expansion:
\begin{equation}
    \boldsymbol{\dot{y}}= \boldsymbol{J}|_{\rm{eq}}(\boldsymbol{r}-\boldsymbol{y})+\boldsymbol{\dot{u}}
\end{equation}
\begin{equation*}
    \boldsymbol{y}=\begin{pmatrix}
           c_1^{\rm{I}} \\[2pt]
           c_2^{\rm{I}} \\[2pt]
           c_1^{\rm{II}} \\[2pt]
           c_2^{\rm{II}} \\[2pt]
           V^{\rm{I}} \\
         \end{pmatrix},\,
         \boldsymbol{r}=\begin{pmatrix}
           c_{1\,\rm{eq}}^{\rm{I}} \\[2pt]
           c_{2\,\rm{eq}}^{\rm{I}} \\[2pt]
           c_{1\,\rm{eq}}^{\rm{II}} \\[2pt]
           c_{2\,\rm{eq}}^{\rm{II}} \\[2pt]
           V_{\rm{eq}}^{\rm{I}} \\
         \end{pmatrix},\,
         \boldsymbol{u}=\begin{pmatrix}
           0 \\[2pt]
           0 \\[2pt]
           u \\[2pt]
           -u \\[2pt]
           0
         \end{pmatrix}
\end{equation*}
$\boldsymbol{J}|_{\rm{eq}}$ is the Jacobian matrix of the system evaluated at phase equilibrium. In order to analyse the frequency response of this system we take the Laplace transform which gives us a system of the form
\begin{align}
    \boldsymbol{\bar{y}}(s)=&(\boldsymbol{I}+\boldsymbol{PC})^{-1}\boldsymbol{PC}\,\boldsymbol{r}+(\boldsymbol{I}+\boldsymbol{PC})^{-1}\boldsymbol{P\,\bar{u}}(s)\,.
\end{align}
The quantities of interest are the transfer function matrix of the inputs to the disturbances 
\begin{equation}
    \boldsymbol{H}_{od}=(\boldsymbol{I}+\boldsymbol{PC})^{-1}\boldsymbol{P}\,.
\end{equation} 
However, as $\boldsymbol{\bar{u}}$ has only two non zero values we get a vector of transfer functions. The transfer functions for the concentration in each phase are
\begin{align}
    H_{\rm{od}}^{\rm{I}}(s) &= \frac{K_g s}{K_d s^2 +K_ps + K_i}\,,\\
    H_{\rm{od}}^{\rm{II}}(s) &= \frac{K_ds^2+K_ls}{K_d s^2+K_ps +K_i}\,.
\end{align}
The transfer functions of each component in each phase are symmetric, so the magnitude and angle of the transfer functions are equivalent. The coefficients depend on equilibrium phase separation properties and are defined as 
\begin{align}
    K_d &= V^{\rm{I}}_{\rm{eq}}V^{\rm{II}}_{\rm{eq}}(c^{\rm{I}}_{\rm{eq}}c^{\rm{II}}_{\rm{eq}})^2\,,\\
    K_p &= -D_e c^{\rm{I}}_{\rm{eq}}c^{\rm{II}}_{\rm{eq}}(1-c^{\rm{I}}_{\rm{eq}}c^{\rm{II}}_{\rm{eq}}(2+\chi))\,,\\
    K_i &= -D_e^2(1-2 c^{\rm{I}}_{\rm{eq}})^2(1+2(c^{\rm{I}}_{\rm{eq}}-1)c^{\rm{I}}_{\rm{eq}}\chi)\,,\\
    K_g&=D_eV^{\rm{II}}_{\rm{eq}}(c^{\rm{I}}_{\rm{eq}}-1)^2(c^{\rm{I}}_{\rm{eq}})^{2}(\chi-2)\,,\\
    K_l&=(V^{\rm{I}}_{\rm{eq}}-1)(c^{\rm{I}}_{\rm{eq}}-1)c^{\rm{I}}_{\rm{eq}}D_e(1-c^{\rm{I}}_{\rm{eq}}(2+\chi)+(c^{\rm{I}}_{\rm{eq}})^{2}(2+\chi))\,.
\end{align}
These coefficients are plotted against the phase separation properties in Fig. S2. We see that $K_p$, $K_g$, and $K_l$ have maxima at specific values for $\chi$. $K_d$ also  has a maxima when $V_{\rm{eq}}^{\rm{I}}=0.5$ but the equilibrium volume does not have an effect on $K_p$ or $K_i$.

\begin{figure}
    \centering
    \includegraphics[width=1\linewidth]{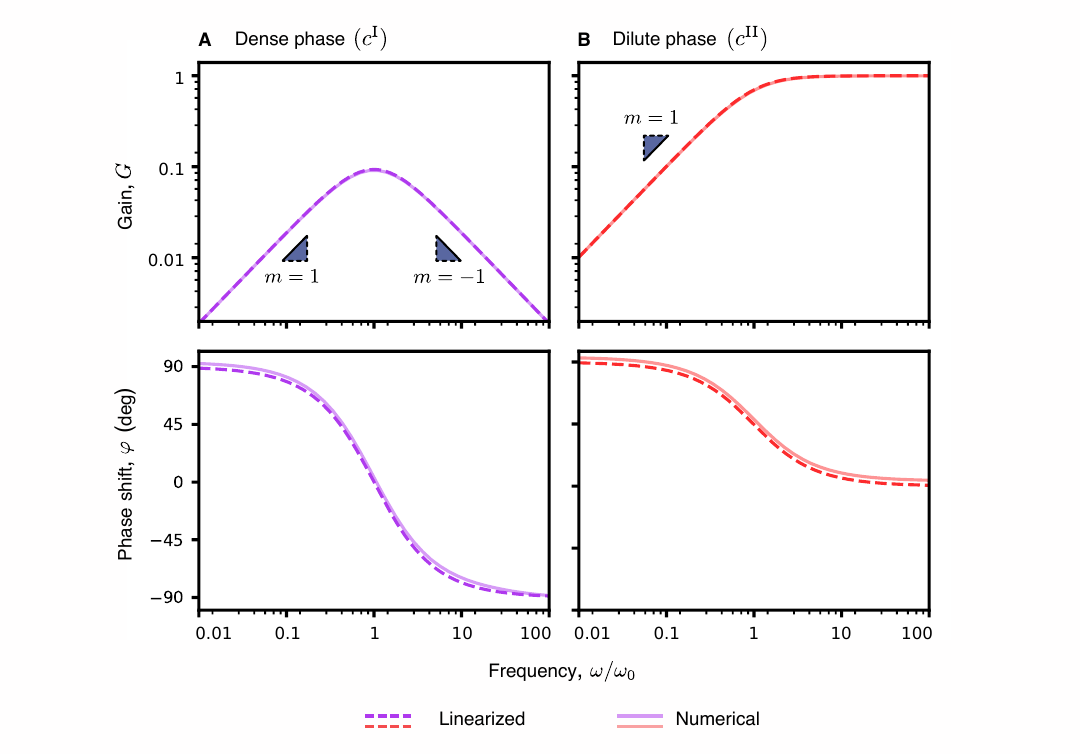}
    \caption{Analytical control-theory predictions and numerical simulations agree, validating our linearized model as a reliable framework for frequency-domain analysis of LLPS dynamics.}
    \label{fig:s1}
\end{figure}
\begin{figure}
    \centering
    \includegraphics[width=0.9\linewidth]{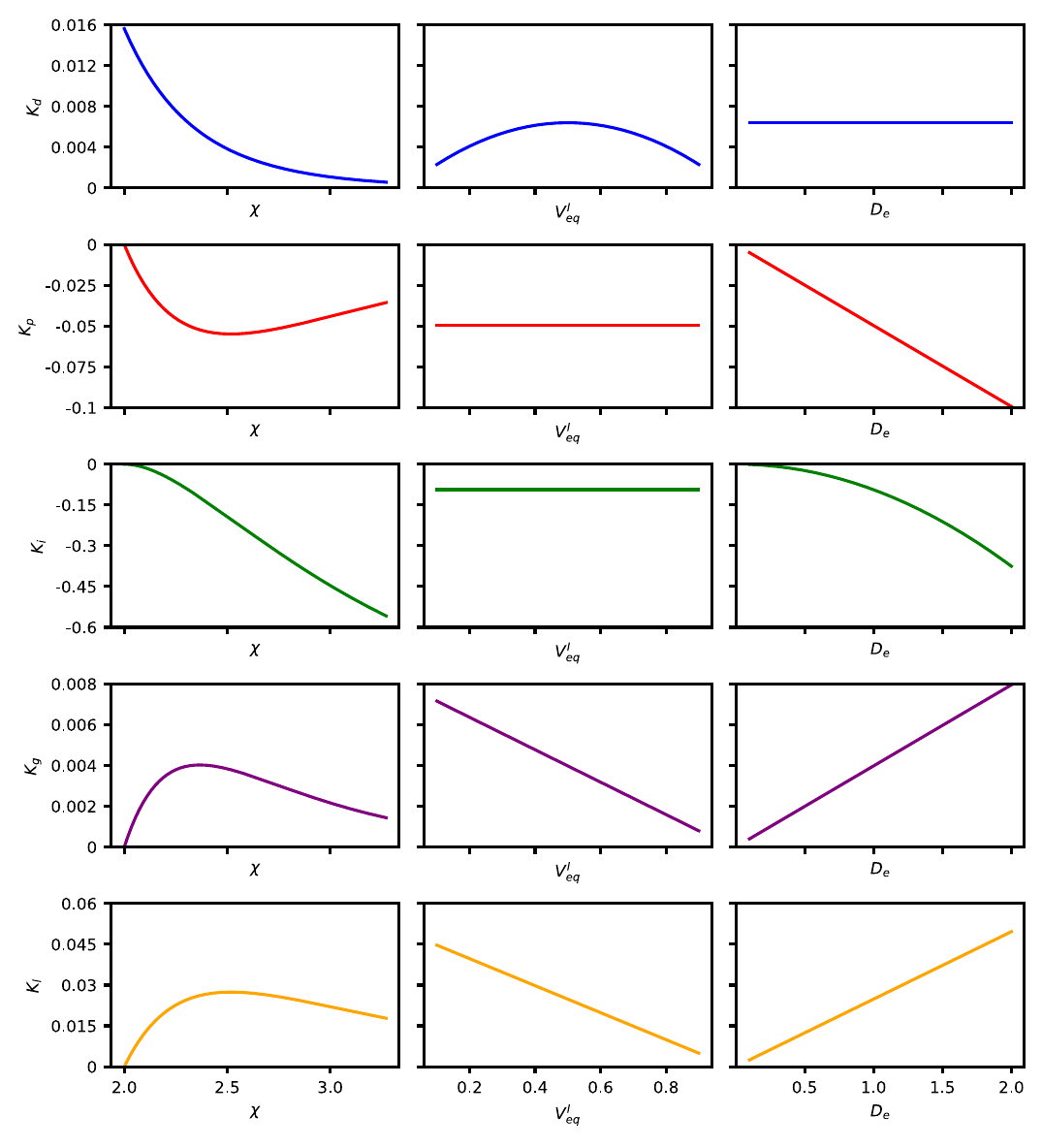}
    \caption{Transfer function coefficients depend systematically on $\chi$, droplet volume, and diffusivity. These dependencies provide the quantitative link between LLPS parameters and dynamic control properties. The parameters were set to $\chi=2.31049$, $V^{\rm{I}}_{\rm{eq}}=0.5$, $D_e = 1$ except for the specific parameter being varied.}
    \label{fig:s2}
\end{figure}
\end{document}